\documentclass[preprint2]{aastex61}
\usepackage[utf8]{inputenc}
\usepackage{rotating}
\usepackage{booktabs}
\usepackage{subfigure}
\usepackage{multirow}

\newcommand{\ha}{H$\alpha$}

\newcommand{\oiii}{[O\thinspace{\sc iii}]}
\newcommand{\oii}{[O\thinspace{\sc ii}]}

\begin{document}

\title{An old stellar population or diffuse nebular continuum emission discovered in green pea galaxies}

\author{Leonardo Clarke}
\affiliation{Minnesota Institute for Astrophysics, University of Minnesota, 116 Church St SE, Minneapolis, MN 55455, USA}

\author{Claudia Scarlata}
\affiliation{Minnesota Institute for Astrophysics, University of Minnesota, 116 Church St SE, Minneapolis, MN 55455, USA}

\author{Vihang Mehta}
\affiliation{Minnesota Institute for Astrophysics, University of Minnesota, 116 Church St SE, Minneapolis, MN 55455, USA}

\author{William C. Keel}
\affiliation{Department of Physics and Astronomy, University of Alabama, Tuscaloosa, AL 35487-0324, USA}

\author{Carolin Cardamone}
\affiliation{Tufts University, 108 Bromfield Road Somerville,  MA  02144, USA}

\author{Matthew Hayes}
\affiliation{Stockholm University, Department of Astronomy and Oskar Klein Centre for Cosmoparticle Physics, AlbaNova University Centre, SE-10691, Stockholm, Sweden.}

\author{Nico Adams}
\affiliation{Minnesota Institute for Astrophysics, University of Minnesota, 116 Church St SE, Minneapolis, MN 55455, USA}

\author{Hugh Dickinson}
\affiliation{School of Physical Sciences, The Open University, Milton Keynes, MK7 6AA, UK}

\author{Lucy Fortson}
\affiliation{Minnesota Institute for Astrophysics, University of Minnesota, 116 Church St SE, Minneapolis, MN 55455, USA}

\author{Sandor Kruk}
\affiliation{European Space Agency (ESA), European Space Research and Technology Centre (ESTEC), Keplerlaan 1, 2201 AZ Noordwijk, The Netherlands} 

\author{Chris Lintott}
\affiliation{Oxford Astrophysics, Department of Physics, University of Oxford, Denys Wilkinson Building, Keble Road, Oxford, OX1 3RH, UK} 

\author{Brooke Simmons}
\affiliation{Department of Physics, Lancaster University, Bailrigg, Lancaster, LA1 4YB, UK} 


\begin{abstract}
We use new HST images of nine  Green Pea Galaxies (GPGs) to study their resolved structure and color.  The choice of filters, F555W and F850LP, together with the redshift of the galaxies ($z\sim 0.25$), minimizes the contribution of the nebular \oiii\ and \ha\ emission lines to the broad-band images. While these galaxies are typically very blue in color, our analysis reveals that it is only the dominant stellar clusters that are blue.   Each GPG does clearly show the presence of at least one bright and compact star-forming region, but these are invariably superimposed on a more extended and lower surface brightness emission.   Moreover,  the colors of the star forming regions are on average bluer than those of the diffuse emission, reaching up to 0.6 magnitudes bluer. Assuming that the diffuse and compact components have constant and single burst Star Formation Histories, respectively, the observed colors imply that the diffuse components (possibly the host galaxy of the star-formation episode) have, on average, old stellar ages ($>1$Gyr), while the star-clusters are younger than 500Myrs. 
We also discuss the possibility that the diffuse red emission is due to a varying relative contribution of nebular continuum, rather than a different stellar population. With the available data, however, it is not possible to distinguish between these two interpretations. A substantial presence of old stars would indicate that the mechanisms that allow large escape fractions in these local galaxies may be different from those at play during the reionization epoch.

\end{abstract}


\section{Introduction}

Green Pea galaxies (GPGs) are low-redshift ($z$) objects first discovered by citizen scientists in the GalaxyZoo project  \citep{lintott2008zoo,cardamone2009galaxy}.  These galaxies turned out to be much more than just an "interesting curiosity". They have low stellar masses  ($\approx$ 10$^{9}M_{\odot}$), high specific star-formation rates ($\approx 10^{-8}$yr$^{-1}$), small sizes (completely unresolved at ground based resolution), and their optical spectra are characterized by  extreme  equivalent-widths (EWs) in the \oiii$\lambda$5007 and \ha\ emission lines \citep[e.g.,][]{Yang2017b,Brunker2020}.  

These galaxies have been the focus of intense research activity since their discovery as they are thought to be among the best known analogs of high-redshift galaxies, and those that were responsible for the reionization of the Universe at $z>6$. Specifically, their compact UV morphologies, low stellar masses, low  metallicities, high specific star-formation rates and high ionizing photon production rate \citep{Cardamone2009,Amorin2010,Izotov2016a} are very similar to those of  typical star-forming galaxies at $z\gtrsim 6$ \citep{Schaerer2016}. 
GPGs are the \emph{only galaxy population} known to have a high escape fraction of hydrogen ionizing radiation, from direct measurement of the stellar continuum below 912\AA, with values ranging between 2-76\% \citep{Izotov2016a,Izotov2016b,Faisst2016,Izotov2018a,Izotov2018b}. The physical conditions that allow these high escape fractions are still poorly understood. One possibility is that these galaxies are overall deficient in neutral hydrogen (e.g., Henry er al. 2015, Eggen et al., in prep), although it is also possible that the ionizing radiation escapes along lines of sight where most of the hydrogen is ionized \citep{Zackrisson2013}. 

Little is known about the star-formation history of GP galaxies, and whether or not they host old stellar populations. The debate about the star-formation history of dwarf starburst galaxies is an old one, with I~Zw~18 as the prototypical example \citep{Aloisi2007,Papaderos2012,Izotov1997}. This galaxy is characterized by a very compact and young starburst, although the detection of red giant branch stars suggests that an older ($\gtrsim 1$Gyr) stellar component also exists. Should the GP galaxies be similar to I~Zw~18, then the mechanisms that drive their high escape fraction may be different than those in the young galaxies during the reionization epoch. 

A limiting factor to detailed studies of their stellar populations is that GPGs's global light  is dominated  by the blinding emission of young stars in the  UV and by the intense emission lines in broad band optical filters. Additionally, most of the results  come from unresolved data from ground-based telescopes (where most observations are seeing-limited). Space based observations have so far been mostly limited to the UV spectral range, using the imaging capabilities of the Cosmic Origin Spectrograph (COS). These studies reveal a UV morphology  characterized by bright star-forming regions in the center of the GPGs, and possibly the presence of exponential discs with scale lengths between 0.6--1.4~kpc \citep[e.g.,][]{Izotov2016a}. These studies, however, are hampered by the limited unvignetted portion of the COS aperture ($\lesssim\,$0\farcs5 radius), and by the fact that they mostly trace the spatial distribution of young stars.

\begin{figure}[hbt!]
\includegraphics[trim={0 0.6cm 0 1.6cm}, width=9cm]{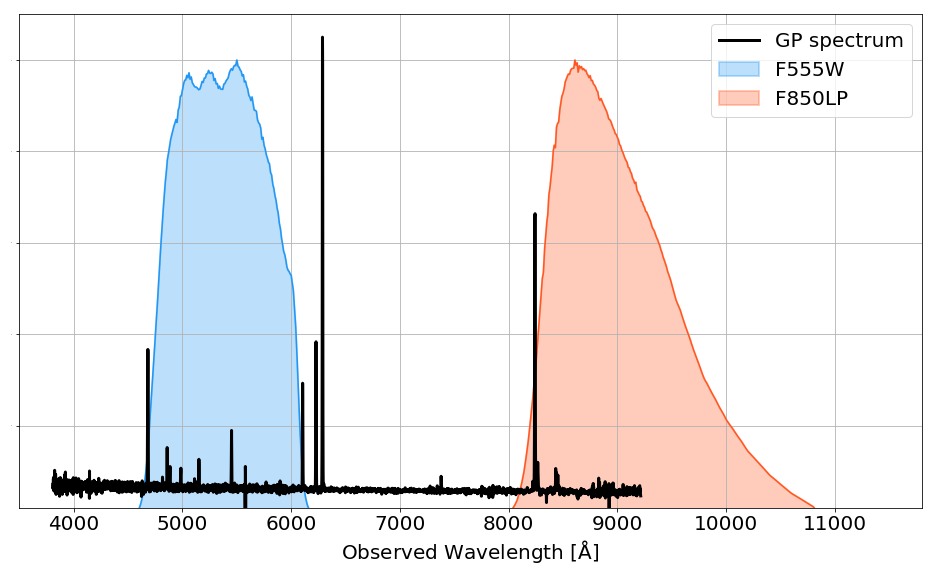}
\caption{Transmission curves of the ACS filters superimposed on the SDSS optical spectrum of J0353--0010. The most prominent emission lines characteristic of the GP spectra are excluded by the filter/redshift combination.\label{fig:strategy}}
\end{figure}

In this study, we analyze new images of nine GPGs taken with the Advanced Camera for Surveys (ACS) on board  the HST. The galaxies were chosen to be at $z\sim 0.25$ -- at this redshift the F555W and F850LP filters exclude the strongest emission lines in  typical GP spectra (e.g., \oiii\ and \ha, see Figure~\ref{fig:strategy}). The high resolution optical imaging allows for a detailed look at the stellar morphology of these galaxies, without contamination from nebular emission that can be substantial in these objects \citep[e.g.,][]{Guseva2017}.  In Section \ref{sec:methods}, we present the data and the analysis. In Section \ref{sec:results} we present our results. Section~\ref{sec:discussion} discusses the color maps and possible physical interpretations of these data. Throughout the paper, magnitudes are expressed in the AB magnitude system \citep{Oke1990}. At $z\sim 0.25$ the physical scale is  4~kpc/$''$, assuming a Planck 2015 cosmology \citep{Planck2015}.

\begin{figure*}[hbt!]
    \centering
     \includegraphics[width=18cm]{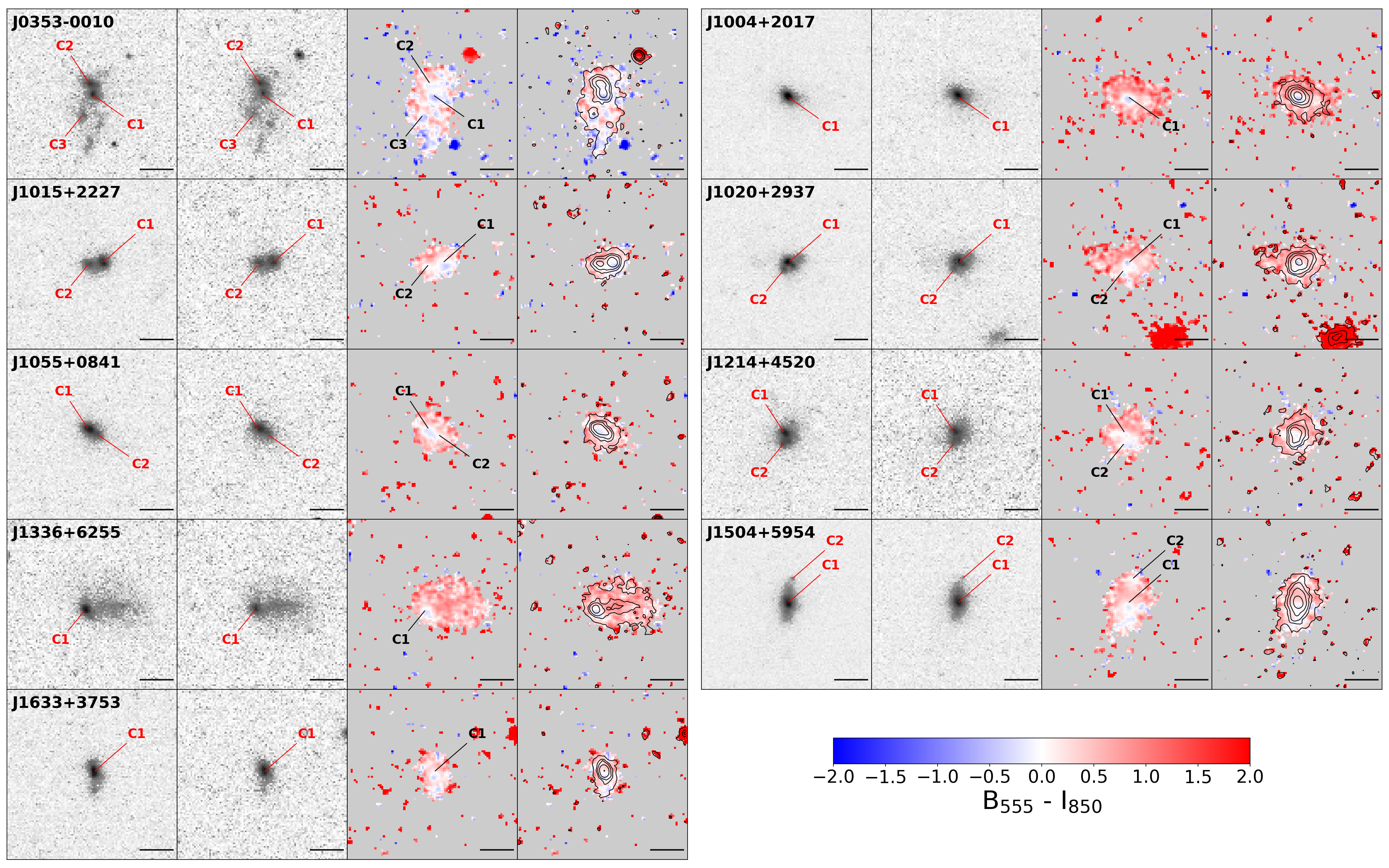}
    \caption{Single band and $B-I$ color maps of the nine target GP galaxies. For each object, we show, starting from the left-most panel: the $B-$ and $I-$band images, the $B-I$ color map with the position of the identified stellar clusters marked, and the $B-I$ color map with the surface brightness contours. The bar in the bottom right corner of each panel indicates 1~arcsecond (corresponding to a physical scale of $\approx$4~kpc at the redshift of the galaxies).}
    \label{fig:images}
\end{figure*}

\begin{deluxetable*}{lccc|ccc}
\tabletypesize{\scriptsize}
\tablecaption{Characteristics of the nine target GPGs. We list each galaxy by name, followed by the stellar clusters detected within each galaxy and their corresponding $B-I$ color. \label{tab:data1}}
\tablehead{\colhead{Target Name} & \colhead{Coordinates} & \colhead{z}  & \colhead{$(B-I)_{\rm G}$} & \colhead{Clump ID} &\colhead{$(B-I)_{\rm ObsSC}$} & \colhead{$(B-I)_{\rm SC}$} \\ 
\colhead{} & \colhead{(J2000)} & \colhead{} & \colhead{(mag)} & \colhead{} & \colhead{(mag)} & \colhead{(mag)}} 
\startdata
\multirow{3}{*}{J0353-0010} & \multirow{3}{*}{03:53:32.4636 -00:10:28.88} & \multirow{3}{*}{0.246} & \multirow{3}{*}{0.06$\pm$0.02} & C1 & -0.15$\pm$0.03 & -0.2$\pm$0.03 \\
                            &                                             &                        &                                & C2 & 0.01$\pm$0.03 & -0.01$\pm$0.03 \\
                            &                                             &                        &                                & C3 & -0.11$\pm$0.07 & -0.29$\pm$0.13 \\
\hline
\multirow{1}{*}{J1004+2017} & \multirow{1}{*}{10:04:00.6406 +20:17:19.25} & \multirow{1}{*}{0.255} & \multirow{1}{*}{0.25$\pm$0.01} & C1 & 0.01$\pm$0.01 & -0.01$\pm$0.01 \\
\hline
\multirow{2}{*}{J1015+2227} & \multirow{2}{*}{10:15:41.1521 +22:27:27.52} & \multirow{2}{*}{0.243} & \multirow{2}{*}{0.16$\pm$0.02} & C1 & -0.1$\pm$0.02 & -0.15$\pm$0.03 \\
                            &                                             &                        &                                & C2 & 0.28$\pm$0.03 & 0.28$\pm$0.04 \\
\hline
\multirow{2}{*}{J1020+2937} & \multirow{2}{*}{10:20:57.4622 +29:37:26.47} & \multirow{2}{*}{0.256} & \multirow{2}{*}{0.21$\pm$0.01} & C1 & -0.05$\pm$0.02 & -0.08$\pm$0.02 \\
                            &                                             &                        &                                & C2 & 0.05$\pm$0.03 & -0.03$\pm$0.03 \\
\hline
\multirow{2}{*}{J1055+0841} & \multirow{2}{*}{10:55:30.4166 +08:41:32.9 } & \multirow{2}{*}{0.252} & \multirow{2}{*}{0.17$\pm$0.02} & C1 & -0.04$\pm$0.02 & -0.07$\pm$0.02 \\
                            &                                             &                        &                                & C2 & 0.13$\pm$0.03 & 0.07$\pm$0.04 \\
\hline
\multirow{2}{*}{J1214+4520} & \multirow{2}{*}{12:14:23.1802 +45:20:40.91} & \multirow{2}{*}{0.255} & \multirow{2}{*}{0.26$\pm$0.02} & C1 & 0.14$\pm$0.02 & 0.1$\pm$0.03 \\
                            &                                             &                        &                                & C2 & -0.01$\pm$0.05 & -0.11$\pm$0.06 \\
\hline
\multirow{1}{*}{J1336+6255} & \multirow{1}{*}{13:36:07.9138 +62:55:30.77} & \multirow{1}{*}{0.252} & \multirow{1}{*}{0.45$\pm$0.02} & C1 & -0.02$\pm$0.03 & -0.09$\pm$0.03 \\
\hline
\multirow{2}{*}{J1504+5954} & \multirow{2}{*}{15:04:57.9874 +59:54:07.27} & \multirow{2}{*}{0.250} & \multirow{2}{*}{0.25$\pm$0.01} & C1 & 0.15$\pm$0.01 & 0.14$\pm$0.01 \\
                            &                                             &                        &                                & C2 & 0.1$\pm$0.07 & -0.4$\pm$0.21 \\
\hline
\multirow{1}{*}{J1633+3753} & \multirow{1}{*}{16:33:37.9414 +37:53:14.3 } & \multirow{1}{*}{0.252} & \multirow{1}{*}{0.2$\pm$0.02} & C1 & 0.17$\pm$0.02 & 0.15$\pm$0.02 \\
\enddata
\end{deluxetable*}

\section{Observations and Analysis}\label{sec:methods}
The targets of this study were selected from the SDSS spectroscopic catalog (Data Release~12), using the same color criteria described in \citet{Cardamone2009}, and limited to star-forming galaxies using the \citet{Kauffmann2003agn} classification,  with the additional redshift constraint of $z=0.250 \pm 0.006$ to ensure that \oii, \oiii, and \ha\ are outside the chosen filters. From the resulting sample, we removed galaxies with bright nearby stars that could compromise the accuracy of the photometry. 

The galaxies were observed with the HST-ACS camera as part of the “Gems of the Galaxy Zoo” (Zoo Gems) program (PID: 15445, PI: Keel). For each galaxy, two 337~s images were taken, one each in the F555W and F850LP filters. At redshift $z=0.250$, the central wavelength of the F555W and F850LP filters correspond to rest frame wavelengths of 4500\AA\ and 7200\AA, respectively. In what follows, we will refer to the F555W and F850LP filters as the $B$ and $I$ band, respectively.
The images were bias-subtracted, dark-corrected, and flat-fielded, using {\it calacs, version 10.2.1}. This new version includes a pixel-based correction for charge transfer efficiency losses that mitigates the amplification of readnoise \citep{ctecorrection}. 
Cosmic ray rejection was performed using a Laplacian edge detection algorithm developed by \citet{van2001cosmic}.
After the cosmic ray removal, we performed sky subtraction on all of the images. The sky value was calculated for each image using the Astrodrizzle\footnote{\url{https://www.stsci.edu/scientific-community/software/drizzlepac.html}} tool.

The images were then photometrically calibrated to AB magnitudes \citep{Oke1990} and corrected for foreground Galactic extinction at the position of each target using the \citet{Schlafly2011} recalibration of the \citet{schlegel1998}  extinction map. We used the NASA/IPAC Extragalactic Database extinction calculator\footnote{\url{https://ned.ipac.caltech.edu/extinction_calculator}} to compute the Galactic extinction in the HST filter bandpasses.

Colors were computed after the homogenization of  the point spread function (PSF) across the two bands. We chose to convolve the F555W images to match the broader PSF of the F850LP band.  First, we generated model PSFs at the position of each target in the detector for each ACS filter using the HST Tiny Tim software \citep{tinytim2011}.  We then generated the F555W-to-F850LP PSF homogenization kernel using Pypher\footnote{\url{https://pypher.readthedocs.io/}} \citep{boucaud16}. Color maps  were then created using the convolved images.

\section{Results}
\label{sec:results}
In Figure~\ref{fig:images} we present the individual images of the  nine GPGs as well as the $B-I$ color maps. All GPGs are clearly resolved, and show a range of morphologies. In most galaxies we identify multiple stellar clusters, likely star-forming regions responsible for the characteristic emission line spectra, in addition to a diffuse and extended continuum. In a few objects, e.g. J1004$+$2017 and J1020$+$2937, we see a single prominent star cluster that lies in the galaxy center, giving rise to a relatively symmetric morphology, while in J1336$+$6255 the cluster is off-centered with respect to the diffuse continuum.

We automatically detected the stellar-clusters within the GPGs using a contrast-based image analysis algorithm, adopted from \citet{Guo2015} and with some modifications \citep{Mehta2020}. 
Briefly,  the $B$ and $I$ stamps were smoothed with a 8px (0.4\arcsec) boxcar kernel, and a contrast image for each filter was generated by subtracting the smoothed image from the original. The contrast image was then filtered to mask out all pixels below 2$\sigma$ (estimated in blank sky regions) to create a filtered image. Finally, the clusters were detected on the $B$+$I$ multi-band filtered images using SExtractor \citep{Bertin1996}. Only those stellar clusters within the SExtractor segmentation map generated from the smoothed $I$ image of the galaxy were retained for further analysis. The positions of the resulting clusters are marked on the $I-$band images in Figure~\ref{fig:images}. Galaxies J0353, J1015, and J1214 host multiple star clusters, while J3336, J1633 and J1004 are examples of objects with only one well defined cluster located either off-center with respect to the underlying continuum (e.g.,  J3336) or close to the galaxy center (e.g., J1004). 

The $B-I$ color maps shown in Figure~\ref{fig:images} reveal that the stellar clusters have colors significantly bluer than the rest of the host galaxy. In order to quantify this difference, we have computed the galaxies' and stellar-clusters'  colors as follows.
First, we used the contrast images to create a segmentation map of each galaxy, with stellar-clusters and host pixels identified. We then computed the host-galaxy color, ($B-I$)$_{\rm G}$, by summing the flux of all pixels not assigned to star-clusters. 
For the stellar clusters, we provide two estimates of their colors, with and without the subtraction of the underlying host-galaxy contribution. To subtract the galaxy contribution, we compute the average host-galaxy surface brightness within the host segmentation map and assume that this value corresponds to the galaxy color at the position of the cluster. Judging by the color maps in Figure~\ref{fig:images}, this assumption appears to be justified as no strong gradients, or variations in colors, are visible within the host galaxies. The local background flux is subtracted from the cluster aperture, and the fluxes of the stellar-clusters and the corresponding colors, ($B-I$)$_{\rm SC}$, were then computed within circular apertures of 0\farcs2 radius. 

In addition to the host-galaxy subtracted cluster colors, we also measured the  colors without removing the galaxy contribution, ($B-I$)$_{\rm ObsSC}$, using the same 0\farcs2 aperture. The colors of the host-galaxy and the stellar clusters are reported in Table~\ref{tab:data1}. 

\begin{figure}
    \centering
    \includegraphics[width=8.5cm]{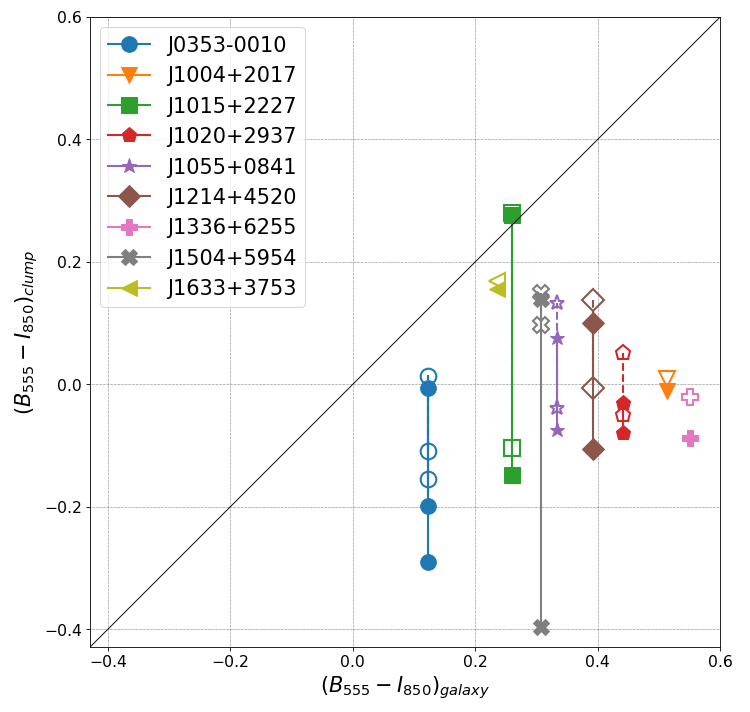}
    \caption{($B-I$) color comparison between each galaxy and the clumps corresponding to each galaxy. Note that points connected by vertical lines correspond to galaxies that host multiple clumps. Solid points shows the colors corrected for the underlying galaxy contribution, while open symbols show the measurement of the observed colors at the position of the clump (see text for details).}
    \label{fig:clump_vs_galaxy_colors}
\end{figure}

In Figure \ref{fig:clump_vs_galaxy_colors}, we present a comparison between the $(B-I)$ colors of the stellar clusters and their corresponding host galaxies. Filled symbols indicate  $(B-I)_{\rm SC}$, while open symbols show the colors observed at the position of the stellar clusters $(B-I)_{\rm ObsSC}$. For galaxies with more than one clump, the stellar-cluster colors are connected by a vertical line.  
Right away, we quantitatively confirm the impression derived from the color maps that the stellar clusters  appear to be bluer than the underlying host galaxy. Some of the clusters are up to 0.6\,magnitudes bluer than the underlying host. This conclusion does not change whether we use the $(B-I)_{\rm SC}$ colors or the $(B-I)_{\rm ObsSC}$ ones, as the galaxy subtracted stellar-cluster colors are only 0.1 magnitudes bluer than the observed colors at the cluster positions.

\begin{figure}
    \centering
    \includegraphics[width=8.5cm]{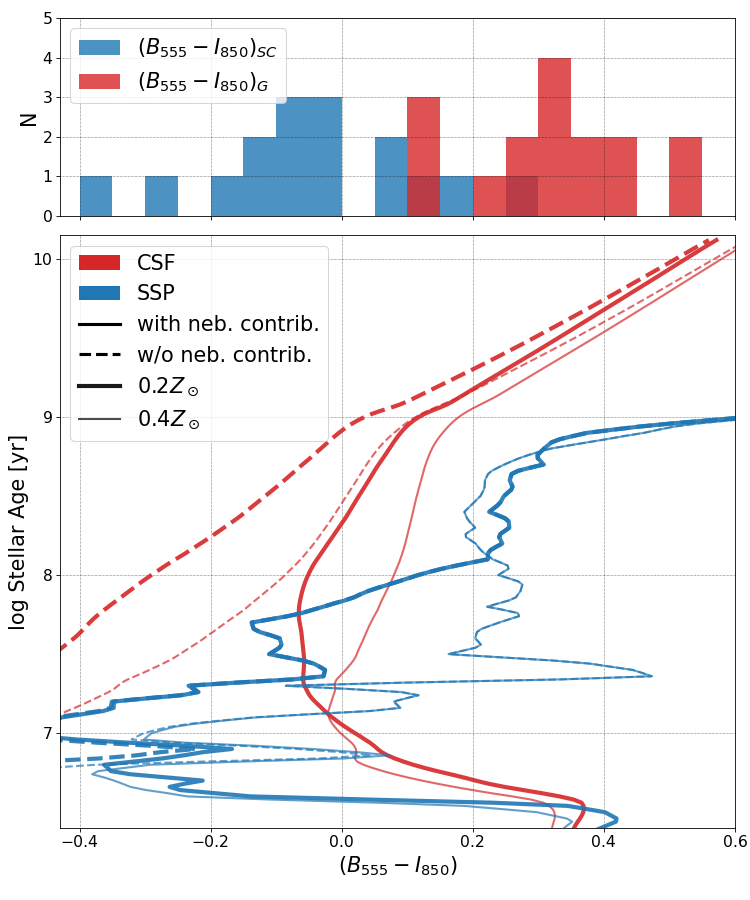}
    \caption{\emph{Top panel:} histogram of galaxy and clump colors.  \emph{Bottom panel:} Stellar age versus $B-I$ color for constant star-formation rate (red) and simple stellar population (blue) star-formation histories of varying metallicities. Solid and dashed lines represent calculations with and without the contribution of nebular emission (see text for details).}
    \label{fig:sfh}
\end{figure}

\section{Discussion}\label{sec:discussion}
The color differences between the stellar clusters and the host galaxy presented in the previous section can be due to a combination of effects, such as stellar age, metallicity, star-formation history as well as dust extinction. With only two pass-bands, however, disentangling these effects of stellar population properties and/or nebular continuum is not possible. Keeping this caveat in mind, however, we can discuss what are the implications of the relative differences in the observed colors, under some empirically motivated assumptions. 

Existing constraints on the stellar population of GPGs come from  the analysis of \textit{spatially unresolved} data, either ground-based optical observations, or far-UV spectra obtained with the HST-COS spectrograph. These data show that the GPGs are characterized by very young stellar ages of only a few million years, low stellar and nebular metallicities and low dust content \citep[e.g.,][]{Guseva2020}. These are, however, luminosity-weighted results, and the derived physical properties are biased by the emission of the brightest youngest stars \citep[e.g.,][]{Chisholm2019}. 
It is reasonable to assume that the star-formation history (SFH) of these clusters is well approximated by a single burst caught at a very young age, given their compact morphology and high EW emission lines. For the underlying galaxy we assume that stars have formed over a more prolonged period of time. We therefore interpret the color differences using two SFHs which we chose to likely bracket the range of expected colors: constant star-formation (CSF) rate  and simple stellar population (SSP).

In Figure~\ref{fig:sfh} we compare the observed $(B-I)$ colors with the colors of synthetic  stellar population synthesis models. The top panel of Figure~\ref{fig:sfh} shows the distribution of colors for the stellar clusters (blue histogram) and galaxies (red histogram). The blue histogram includes a larger number of objects than the red histogram because some galaxies host multiple clumps.  For each SFH, the theoretical colors are computed for a range of ages (between a few Myrs and the age of the Universe) and two metallicities (20 and 40\% solar). Subsolar metallicities in GPGs are implied both by the analysis of the nebular lines in the optical spectra \citep[e.g., ][]{Yang2017,Izotov2016a} and by the modeling of the FUV spectra observed with COS.  The model templates were generated using Flexible Stellar Population Synthesis  with a \citet{chabrier2003} initial mass function and the default MILES stellar library with MIST isochrones \citep[FSPS][]{Conroy2009,Conroy2010}. We show only dust free models. The effect of an absorption of $A_V=1$  would be to redden the $B-I$ colors by $\approx 0.5$ magnitudes. We show calculations with and without the nebular contribution (from both emission lines and continuum), as solid and dashed lines, respectively. Although the emission lines fall mostly outside the filters, the nebular continuum would still contribute substantially to the observed colors, particularly for younger ages. As Figure~\ref{fig:sfh} shows, for the CSF rate model (red curves) the nebular continuum is important for ages up to one billion years, and dominates  for ages below a few hundred million years. 

Comparing the host-galaxy and stellar cluster colors with CSF rate and SSP models, respectively, we can draw the following conclusions. Neglecting the reddening due to dust extinction, the colors of the host galaxies are well reproduced by stellar populations that have formed continuously over at least the past billion year, up to 10 billion years, or by a burst that occurred more than 100 Myr in the past. This conclusion does not substantially depend on the metallicity of the stars, nor the contribution of the nebular continuum that, at these ages, is subdominant compared to the stellar continuum. The bluer galaxies can, however, also be reproduced with young CSF rate models (younger than 10Myrs) if the nebular continuum is important. The colors of the compact star-forming  regions, on the other hand, suggest younger stellar ages than the underlying galaxies. Specifically, with the single-burst assumption the observed colors can be explained with ages between one and 10Myrs, although ages of a few hundred million years cannot be ruled out. The younger estimates are more in agreement with the previous  studies of GPGs mentioned above. 

The stellar mass hiding behind the blinding stellar clusters is possibly not negligible, although we refrain from estimating its possible contribution given the large uncertainties in the stellar population properties implied by the limited information available. 

Finally, we consider the alternative possibility that there is only one young stellar component, but the young stars and the ionized gas have different spatial distribution \citep[as observed, e.g., in I~Zw~18 by][]{Papaderos2012}. In this scenario, the observed color maps would be the result of a varying relative contribution of the nebular continuum to the total color in different places of the ionized nebula.   

With our choice of filters, the nebular continuum has a red $B-I$ color, of $\approx$0.76, and will only get redder in the presence of dust. If this interpretation is correct, the blue emission at the position of the stellar clusters would be dominated by stellar continuum, with minimal contribution from the ionized gas. This would be possible if, for example, we were looking at the stellar clusters through clear channels in the gas distribution. The emission in the diffuse component of the GPGs would then result from a declining surface brightness in hot stars, and a correspondingly relatively-more-important contribution of the nebular continuum. This scenario is supported by far UV observations of similar galaxies, showing that 
GPGs host extended exponential discs with lower surface brightness than the bright star--forming regions 
\citet{Izotov2016b}. 

We proceed to calculate the expected contribution of nebular continuum over the segmented area of the image.  We computed the predicted nebular continuum flux density from the observed \ha\ flux, assuming an \ha\ $EW_{\rm neb}=4740$\AA, appropriate for a pure nebular spectrum. With this assumption, $f^{\rm neb}_{\lambda}=1.1\frac{f_{\rm H\alpha}}{EW_{\rm neb}}$ (the 10\% correction is applied to convert the continuum from 6562\AA\ to 8750\AA). We use the galaxy half-light radius measured in the $I-$band images (R$_{1/2}$) and compute the expected surface brightness due to the nebular continuum as $I_{\rm neb}=0.5\frac{f_{\rm neb}}{\pi R_{1/2}^2}$. This calculation shows that for a flat surface brightness profile in the nebular gas, the nebular continuum reaches a comparable surface brightness to the diffuse $I-$band emission that is sufficient in each case, and gives $\approx$ four times as many photons as needed in the highest extreme.  More realistic nebular surface brightness profiles will decline with radius, leading to lower contributions.

\section{Conclusion}
We analyzed HST images of nine compact, star-forming Green Pea galaxies at $z\sim 0.25$. The choice of broad band filters avoids the prominent emission lines from these galaxies and allows us to measure resolved colors for the first time. In all galaxies, we find one or more dense bright stellar clusters (typically located  in the central regions of the galaxies) superimposed on a more diffuse component. The bright clumps are associated with  blue colors, systematically bluer than the diffuse component. 

Interpreting the color differences depends on assumptions on the star-formation history in each component, a parameter that is poorly constrained in this class of galaxies. 
Assuming that the diffuse and compact components have constant and single burst SFHs, respectively, the observed colors imply that the diffuse components (possibly the host galaxy of the star-formation episode) have, on average, old stellar ages ($>1$Gyr), while the star-clusters are younger than 500Myrs. 
We discuss the possibility that the diffuse red emission is due to a varying relative contribution of nebular continuum, rather than a different stellar population. With the available data, however, it is not possible to distinguish between these two interpretations.
Green Pea galaxies are the focus of a large investment of resources because they are the only galaxy population for which hydrogen ionizing radiation is consistently observed to escape in large fraction. As such, these objects are often referred to as the best local analogs of the sources that reionized the Universe at $z>6$.  Understanding the properties of these objects will shed light on the mechanisms that drive this high escape fraction. A substantial presence of old stars would indicate that the mechanisms that allow large escape fractions in these local galaxies may be different than those at play during the reionization. Moreover, the different integrated stellar masses, which dominate the gravitational potential in the central regions, may influence the formation of galaxy winds and metal retention.
If confirmed, our result would imply that GPGs are not real analogs of EoR objects, which could be at most few hundreds of million years old at $z > 7$.

\acknowledgments 
This research is partially supported by the National Science Foundation under grant AST 1716602. Based on observations made with the NASA/ESA Hubble Space Telescope. STScI is operated by the Association of Universities for Research in Astronomy, Inc., under NASA contract NAS 5-26555. 
This research has made use of the NASA/IPAC Extragalactic Database (NED), which is funded by the National Aeronautics and Space Administration and operated by the California Institute of Technology.

\facility{HST}

\software{astropy\footnote{http://www.astropy.org} a community-developed core Python package for Astronomy \citep{astropy:2013, astropy:2018}
}

\bibliography{Clarke_v1}
\bibliographystyle{aasjournal}
\end{document}